\documentclass[
reprint,
superscriptaddress,
nofootinbib,
amsmath,amssymb,
aps,
prl,
floatfix,
]{revtex4-1}

\usepackage{graphicx}
\usepackage{dcolumn}
\usepackage{bm}
\usepackage{epstopdf}
\usepackage{braket}

\begin{document}
\title{Coherent Control of Trapped Ion Qubits with Localized Electric Fields}
\author{R. Srinivas}
\email{r.srinivas@oxionics.com}
\affiliation{Oxford Ionics, Oxford, OX5 1PF, UK}
\affiliation{Department of Physics, University of Oxford, Clarendon Laboratory, Parks Road, Oxford, OX1 3PU, UK}
\author{C. M. L{\"o}schnauer}
\affiliation{Oxford Ionics, Oxford, OX5 1PF, UK}
\author{M. Malinowski}
\affiliation{Oxford Ionics, Oxford, OX5 1PF, UK}
\author{A. C. Hughes}
\affiliation{Oxford Ionics, Oxford, OX5 1PF, UK}
\author{R. Nourshargh}
\affiliation{Oxford Ionics, Oxford, OX5 1PF, UK}
\author{{V.~Negnevitsky}}
\affiliation{Oxford Ionics, Oxford, OX5 1PF, UK}
\author{D. T. C. Allcock}
\affiliation{Oxford Ionics, Oxford, OX5 1PF, UK}
\affiliation{Department of Physics, University of Oregon, Eugene, OR 97403}
\author{S. A. King}
\affiliation{Oxford Ionics, Oxford, OX5 1PF, UK}
\author{C. Matthiesen}
\affiliation{Oxford Ionics, Oxford, OX5 1PF, UK}
\author{T. P. Harty}
\affiliation{Oxford Ionics, Oxford, OX5 1PF, UK}
\author{C. J. Ballance}
\affiliation{Oxford Ionics, Oxford, OX5 1PF, UK}
\affiliation{Department of Physics, University of Oxford, Clarendon Laboratory, Parks Road, Oxford, OX1 3PU, UK}

\begin{abstract}
We present a new method for coherent control of trapped ion qubits in separate interaction regions of a multi-zone trap by simultaneously applying an electric field and a spin-dependent gradient. Both the phase and amplitude of the effective single-qubit rotation depend on the electric field, which can be localised to each zone. We demonstrate this interaction on a single ion using both laser-based and magnetic field gradients in a surface-electrode ion trap, and measure the localisation of the electric field.
\end{abstract}

\maketitle

Trapped ion systems are a leading platform for quantum computation due to their excellent coherence properties~\cite{Langer2005, Wang2021}, high-fidelity single~\cite{Harty2014} and two-qubit~\cite{Ballance2016, Gaebler2016, Srinivas2021, Clark2021} operations. A promising route towards a larger-scale trapped-ion quantum processor is the `quantum charge-coupled device' (QCCD) architecture~\cite{Wineland1998,Kielpinski2002}, where ions are stored in separate interaction zones. The ions are transported using time-dependent electric potentials applied to neighbouring electrodes, for example to bring them to specific zones for individually addressed single-qubit operations, as required for universal computation~\cite{Barenco1995}. To minimise the overall computation duration, these individually addressed operations can be performed in parallel~\cite{Aharonov1996}, which requires local control of the phase and amplitude of the qubit drive field resonant with the qubit frequency.

Local single-qubit operations are commonly performed using lasers, focused on individual ions~\cite{Naegerl1999}, or routed to each zone via integrated optics~\cite{Niffenegger2020, Mehta2020}. As the wavelengths of the lasers required ($\sim 500$\,nm) are typically much smaller than the distances between trapping zones ($\sim 0.1-1$\,mm), crosstalk can be minimised. While laser-free methods using magnetic fields have been used for global single-qubit rotations~\cite{Brown2011}, localising this field to individual zones or ions is a challenge due to the long wavelengths of the microwave and radiofrequency fields required ($\sim 1$\,mm$-1$\,m). Instead, individual ions in a global field can be addressed by separating them in frequency space. Confining the ions close to a current-carrying electrode~\cite{Wineland1998, Ospelkaus2008} or a permanent magnet~\cite{Mintert2001} can achieve significant spatial variation of the magnetic field over typical ion separations ($\sim1-10$ \textmu m). Static magnetic field gradients can create a differential Zeeman shift~\cite{Mintert2001, Piltz2014}, or oscillating magnetic-field gradients can induce differential ac Zeeman shifts on the qubits~\cite{Warring2013b, Srinivas2021}. Multiple tones can also be used to coherently cancel magnetic fields at some ions but not others~\cite{Warring2013b, Craik2017}.

\begin{figure}[h]
\centering
{\includegraphics[width=\columnwidth]{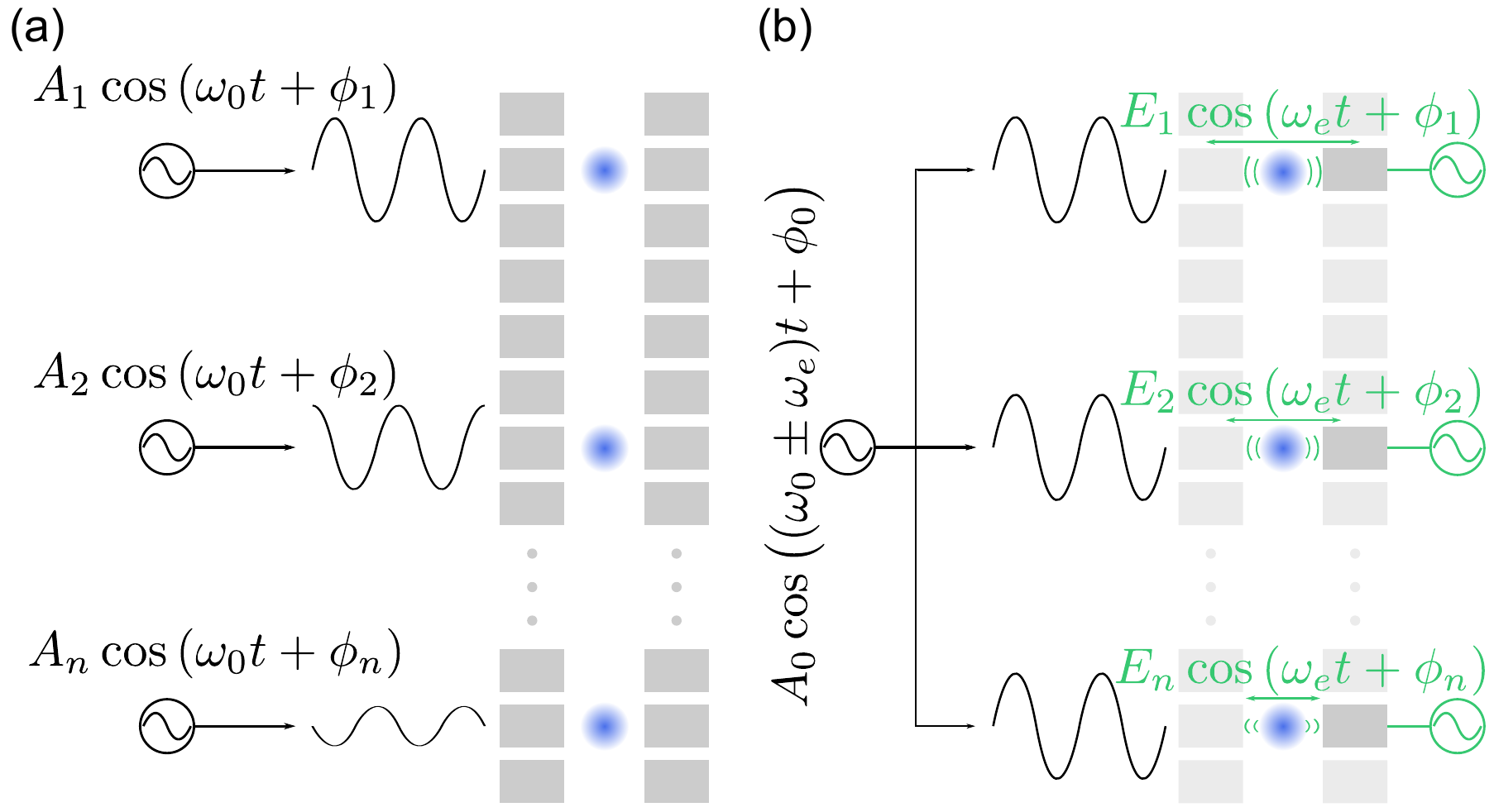}}
\caption{\label{fig_schematic} Fully parallel qubit control in a multi-zone trapped-ion processor. Ions (blue) with qubit frequency $\omega_0$ are trapped in multiple zones, where the confining potential in one dimension is generated by static voltages on dc electrodes (shown in grey). (a)  Existing methods achieve fully parallel control by locally modulating the qubit drive amplitude $A_i$ and phase $\phi_i$. (b) With our scheme, fully parallel control is achieved by local electric fields oscillating at frequency $\omega_e$ with amplitude $E_i$ and phase $\phi_i$. These fields drive the ion motion and, when combined with a single source that drives a global gradient of fixed amplitude and phase at frequency $\omega_0 + \omega_e$ or $\omega_0 - \omega_e$, enable local single-qubit control. The electric field can be generated by applying an oscillating voltage to a single dc electrode per zone.}
\end{figure}

For each of these methods, qubit control in each zone is accomplished by modulating the amplitude and phase of the driving laser or magnetic field (see Fig.~\ref{fig_schematic}a). When scaling to large devices, these modulators would ideally be integrated into the trap to minimise the number of separate signals that need to be routed to the chip. This integration significantly increases the complexity of the device and its fabrication~\cite{Dong2022}. A control method that leverages the already existing infrastructure for the delivery of static or oscillating voltages to trap electrodes would vastly simplify this problem.

Instead of modulating the qubit drive, such voltages applied to the trap electrodes can generate electric fields that modulate the ion motion. For example, when the ion is shifted off the radiofrequency null in a Paul trap, the electric field at the trapping frequency drives micromotion which modulates the fields seen by the ion~\cite{Berkeland1998}. This micromotion can be used for addressing using lasers~\cite{Leibfried1999}, or in combination with a magnetic field gradient~\cite{Warring2013a, Warring2013b}. However, as there is usually only one trap drive that defines the micromotion frequency and phase, there is no local phase control, and any uncompensated electric fields that move the ion off the radiofrequency null can lead to crosstalk. There has also been a recent proposal of using an electric field in combination with spin-dependent forces for addressing~\cite{Sutherland2022}, but the phase of the interaction is governed by the global gradient rather than the local electric field.

Here, we present a new method of performing single-qubit rotations using a spin-dependent gradient in combination with an applied electric field. The phase and amplitude of the operation can be controlled by the phase and amplitude of the electric field, enabling control of multiple zones in parallel, where each zone has its own local electric field as shown in Fig.~\ref{fig_schematic}b. The frequency of the electric field can also be used to compensate for any qubit frequency shifts between different zones. Further, even relatively small oscillating voltages applied to existing electrodes can generate the required ion motion, simplifying control integration and scaling. We demonstrate and characterise this method using a single ion in a surface-electrode ion trap and measure the localisation of the electric field.

We apply an oscillating gradient and an electric field to a single ion, generating the interaction

\begin{align}
\label{eq_bare}
\hat{H} = &\hbar\Omega_g \hat{\sigma}_i\cos{\omega_g t}(\hat{a}+\hat{a}^\dagger) \\ \nonumber 
+&\hbar\Omega_e \cos{(\omega_e t+\phi_e)}(\hat{a}+\hat{a}^\dagger),
\end{align}

\noindent where the first term corresponds to a spin-dependent gradient with coupling strength $\Omega_g$ oscillating at a frequency $\omega_g$. This gradient couples the internal states of an ion to its motion via the Pauli spin operator $\hat{\sigma}_i$, where $i\in\{x, y, z\}$, and the creation (annihilation) operator $\hat{a}^\dagger$ $(\hat{a})$. The second term describes the effect of the electric field with coupling strength $\Omega_e\equiv {qEr_0}/{\hbar}$, where the ion charge is $q$, the electric field amplitude along the motional mode is $E$, $r_0$ is the ground state extent of the ion motion, and $\hbar$ is the reduced Planck constant. This electric field is oscillating at frequency $\omega_e$ with a controllable phase $\phi_e$ relative to the gradient term. In contrast to the spin-dependent gradient term, the electric field exerts a spin-independent force on the ion. 

\begin{figure}[!t]
\centering
{\includegraphics[width=\columnwidth]{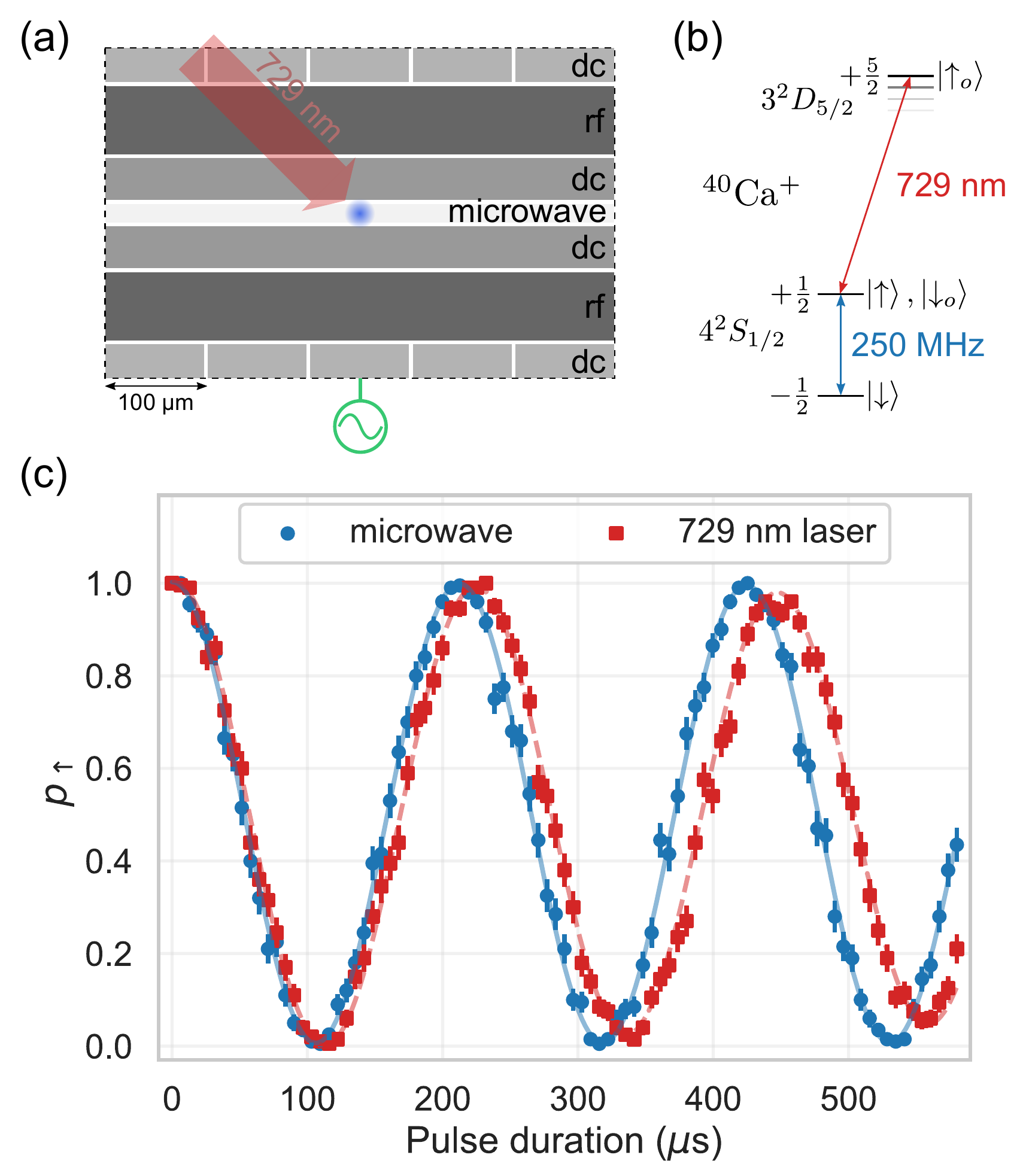}}
\caption{\label{fig_flop}Experimental implementation. (a) Top view of surface-electrode trap section. We use a surface-electrode trap with a single $^{40}$Ca$^{+}$ ion (blue, not to scale) trapped approximately 80\,\textmu m above the surface. We apply an oscillating voltage to one of the outer dc electrodes to generate an oscillating electric field at the ion. The forced motion sideband also requires a spin-dependent gradient; we additionally apply a current to the integrated microwave electrode to create a magnetic field gradient, or use a 729\,nm laser beam. Additional dc electrodes on either side of the trap are omitted. (b) Simplified level diagram of $^{40}$Ca$^+$. We use either the $\ket{\downarrow},\ket{\uparrow}$ Zeeman qubit, or the $\ket{\downarrow_o},\ket{\uparrow_o}$ optical qubit. (c) Rabi flopping using forced motion sidebands with either a microwave gradient (blue circles) or a 729\,nm laser beam (red squares). The ion is initialised in the $\ket{\uparrow}=\ket{\downarrow_o}$ state. We plot the population of the $\ket{\uparrow}$ state versus the pulse duration of the forced motion interaction. Lines are sinusoidal fits with an exponential decay. Error bars indicate 68\% confidence intervals.}
\end{figure}

\begin{figure*}[!t]
\centering
{\includegraphics[width=\textwidth]{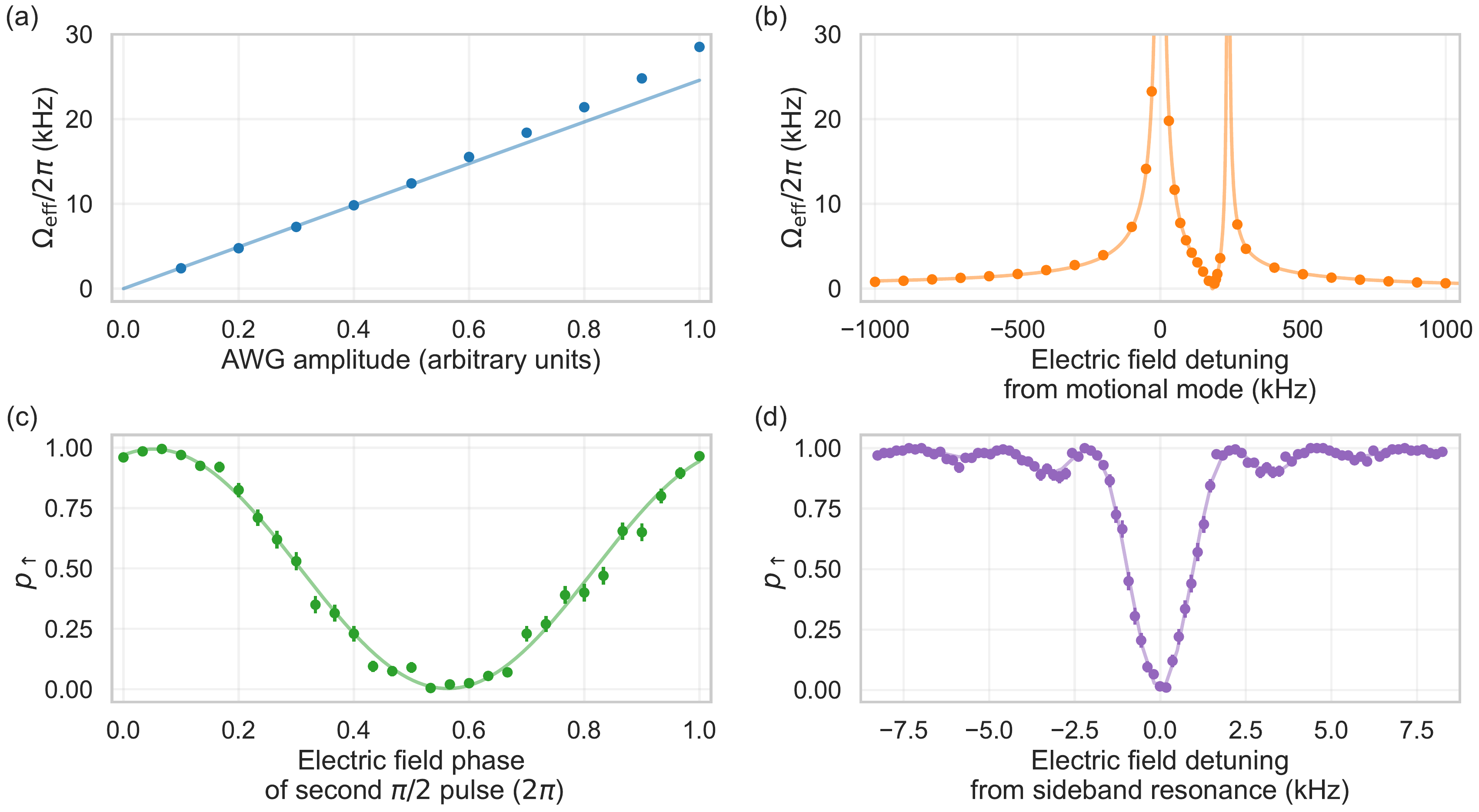}}
\caption{\label{fig_local} Local control of the forced motion sideband with the electric field. (a) The effective Rabi frequency $\Omega_\textrm{eff}$ versus the output amplitude of the AWG used to generate the oscillating voltage, with $(\omega_e-\omega_y)/2\pi=$-100\,kHz. The blue line is a linear fit to the first five data points, following Eq.~(\ref{eq_sf}). The discrepancy at larger amplitudes is likely due to nonlinearities in the signal chain from the AWG. (b) $\Omega_{\textrm{eff}}$ versus the detuning of the electric field frequency $\omega_e$ from the motional modes. Using an electric field amplitude of $E=1.8\,$V/m, we vary $\omega_e$, while simultaneously adjusting the microwave gradient frequency to $\omega_0+\omega_e$. At zero detuning, $\omega_e/2\pi=\omega_y/2\pi=2.6$\,MHz. The second resonance corresponds to the other radial mode at $\omega_z/2\pi=2.8$\,MHz, for which the electric field projection is lower. The orange line is a fit based on Eq.~(\ref{eq_sf}), but includes contributions from both radial modes.  For the plots in (a) and (b), we determine $\Omega_\textrm{eff}$ from Rabi flopping data (e.g. Fig.~\ref{fig_flop}(c)). (c) Dependence of the forced motion sideband phase on the electric field phase $\phi_e$. Starting in the $\ket{\uparrow}$ state, we perform two $\pi/2$ pulses using the forced motion sideband and vary $\phi_e$ in the second $\pi/2$ pulse. We plot the population in the $\ket{\uparrow}$ state versus $\phi_e$ and the green line is a sinusoidal fit. For these data, $E=1.8\,$V/m and $(\omega_e-\omega_z)/2\pi=100\,$kHz. (d) Population remaining in the initial $\ket{\uparrow}$ state versus electric field frequency $\omega_e$. For these data, we keep the microwave gradient at a fixed detuning of 2.1\,MHz from the qubit frequency. When $\omega_e/2\pi=2.1$\,MHz, we drive spin-flips resonantly. We fit a Rabi lineshape to the data~\cite{Rabi1937}. Error bars indicate 68\% confidence intervals.}
\end{figure*}

To evaluate the dynamics, we first go into the interaction picture with respect to the bare Hamiltonian ${\hat{H}_0=\hbar\omega_0\hat{\sigma}_z/2+\hbar\omega_m\hat{a}^\dagger\hat{a}}$, where $\omega_0$ and $\omega_m$ are the qubit and motional frequencies, respectively. We further transform into the interaction picture with respect to the electric field term in Eq.~(\ref{eq_bare}). To perform single-qubit rotations, $i\in\{x, y\}$, and the gradient frequency should be ${\omega_g=\omega_0\pm\omega_e}$. For $i=x$, the interaction Hamiltonian, after dropping faster-rotating terms, is 

\begin{align}
\label{eq_sf}
\hat{H}_\textrm{eff} = \frac{\hbar\Omega_g\Omega_e\omega_m}{2(\omega_e^2-\omega_m^2)}(\cos{\phi_e}\hat{\sigma}_x\mp\sin{\phi_e}\hat{\sigma}_y),
\end{align}

\noindent which drives single-qubit rotations with an effective Rabi frequency ${\Omega_\textrm{eff}\equiv\Omega_g\Omega_e\omega_m/(\omega_e^2-\omega_m^2)}$ and phase $\mp \phi_e$~\cite{Srinivas2020, supplementary}. This interaction, henceforth referred to as the forced motion sideband, can be viewed as involving an off-resonant driving of the mechanical motion of the ion due to the E-field with an amplitude that is proportional to $1/(\omega_e^2-\omega_m^2)$. As the gradient couples the spin to the motion, the driven motion can lead to effective spin-flips when the gradient is detuned from the qubit frequency by the electric field frequency. Crucially, the phase and amplitude of the forced motion sideband are both controllable with the applied (local) electric field.

We implement this scheme in a cryogenic surface-electrode trap at 5 K. We trap a single $^{40}$Ca$^+$ ion approximately 80\,\textmu m above the chip surface while applying a static magnetic field of $\approx$\,9 mT. The information is encoded in either the Zeeman qubit ${\ket{\downarrow}\equiv4{^2}S_{1/2}\ket{m_J=-1/2}}$, ${\ket{\uparrow}\equiv4{^2}S_{1/2}\ket{m_J=+1/2}}$, or the optical qubit ${\ket{\downarrow_o}\equiv4{^2}S_{1/2}\ket{m_J=+1/2}}$, ${\ket{\uparrow_o}\equiv3{^2}D_{5/2}\ket{m_J=+5/2}}$. We use the field from an integrated microwave electrode to manipulate the Zeeman qubit ($\omega_0/2\pi\approx 250$\,MHz), and a 729\,nm laser to drive the optical transition as shown in Fig.~\ref{fig_flop}(b). The motional mode frequencies are $(\omega_x, \omega_y, \omega_z)/2\pi\approx(1.0, 2.6, 2.8)$\,MHz. The axial mode frequency is $\omega_x$, while $\omega_y$ and $\omega_z$ denote the frequencies of the two radial modes, which are oriented approximately $45^\circ$ to the trap surface. For these experiments, we cool all the motional modes to close to the ground state.

The forced motion sideband requires both a spin-dependent gradient and an electric field. For the spin-dependent gradient, we apply a microwave gradient of $\sim3$\,T/m along the radial motional modes or $\sim0.5$\,mW of the 729 laser using a $1/e^2$ beam radius of 25 \textmu m. These values correspond to $\Omega_g/2\pi=0.5$\,kHz or 1\,kHz for the microwave and laser, respectively. We drive the forced motion sideband at frequency $\omega_g=\omega_0+\omega_e$. We create the electric field by applying an oscillating voltage from an arbitrary waveform generator (AWG) to a nearby dc electrode, as shown in Fig.~\ref{fig_flop}(a). As the electric field has a frequency close to the motional mode, we smoothly ramp the voltage on and off to minimise any residual motional excitation after the pulse~\cite{supplementary}. We employ square pulses for the spin-dependent gradients, either from the 729\,nm laser or the microwaves, that turn on after the electric field has finished ramping. 

First, we demonstrate Rabi flopping via the forced motion sideband using either the microwave gradient or the 729 laser as shown in Figure~\ref{fig_flop}(c). For these data, we apply an oscillating electric field with frequency $\omega_e/2\pi=2.5\,$MHz, -100 kHz detuned from the radial mode at $2.6\,$MHz. With the microwave gradient, we use an electric field of $E=1.2$\,V/m, as estimated from independent measurements on the lower-frequency radial mode~\cite{supplementary}. This electric field corresponds to an oscillating voltage with amplitude 3\,mV on the dc electrode. With the 729 laser, we set the electric field to 0.6 V/m, such that the forced motion sideband has an effective Rabi frequency $\Omega_\textrm{eff}/2\pi \approx5$\,kHz for both methods. We observe a decay in the contrast of the Rabi oscillations when using the 729\,nm beam, likely due to phase noise on the laser.

We characterise the dependence of the forced motion sideband interaction on the electric field parameters in Fig.~\ref{fig_local}. For the data in this figure, we only use the interaction with the microwave gradient. Fig.~\ref{fig_local}(a) and (b) show that the forced motion sideband Rabi frequency $\Omega_\textrm{eff}$ can be set by the amplitude of the electric field, or the detuning of its frequency $\omega_e$ from the motional modes. In addition, the electric field phase $\phi_e$ sets the phase of the forced motion sideband as seen in Fig.~\ref{fig_local}(c). The frequency of the electric field $\omega_e$ can compensate for any differences in the qubit frequency in different zones. Keeping the gradient frequency $\omega_g$ fixed, we can vary $\omega_e$ to drive the forced motion sideband resonantly when $\omega_g-\omega_e=\omega_0$  (see Fig.~\ref{fig_local}(d)).

\begin{figure}[!t]
\centering
{\includegraphics[width=\columnwidth]{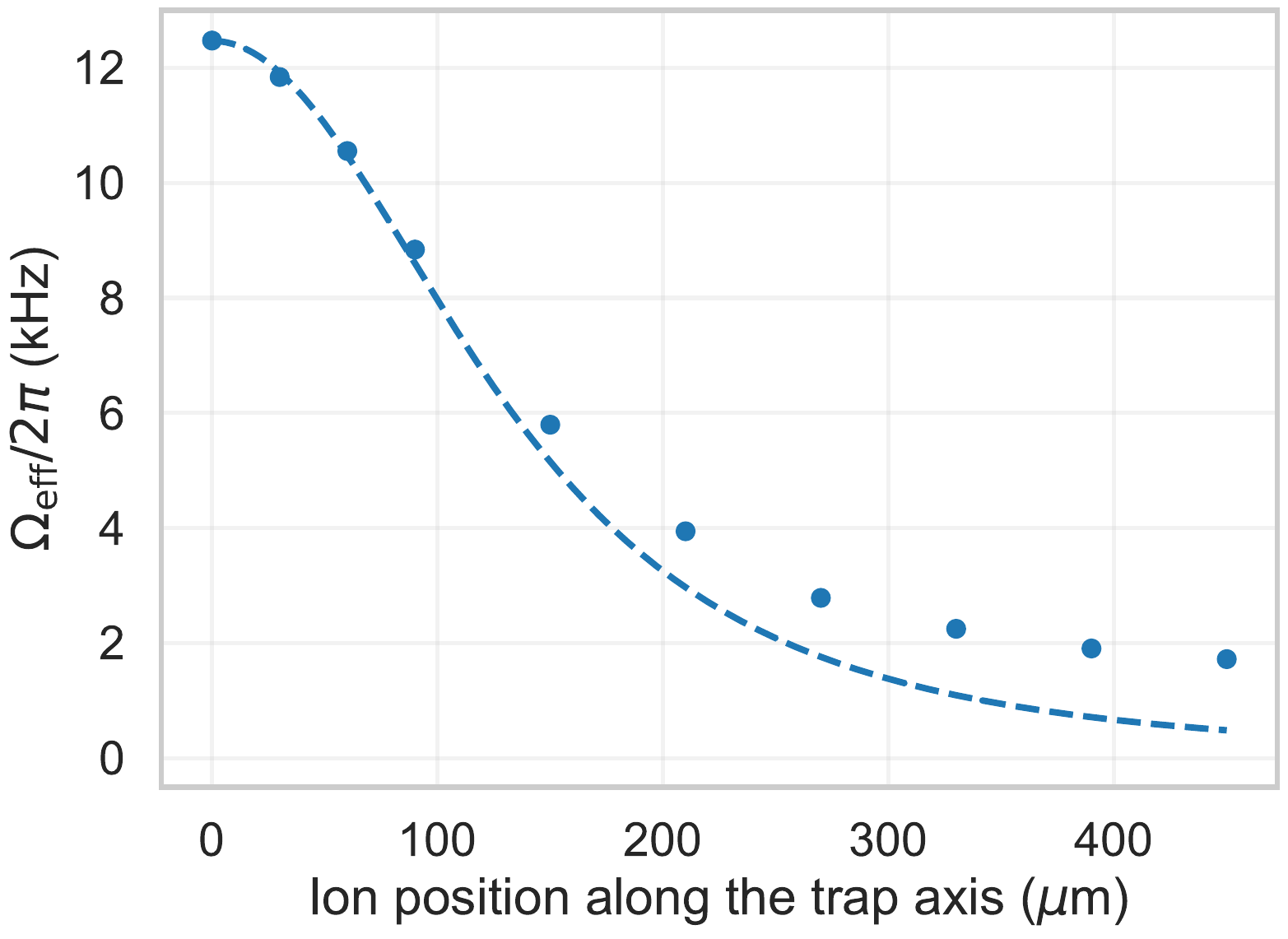}}
\caption{\label{fig_e_field} Localisation of the electric field. We plot the effective spin-flip Rabi frequency $\Omega_\textrm{eff}$ versus the position of the ion along the trap axis. For these data, we use a microwave gradient in addition to the electric field to generate the forced motion sideband interaction. The electric field amplitude is approximately $3\,$V/m with $(\omega_e-\omega_y)/2\pi=-100$\,kHz. At 0\,\textmu m, the ion is aligned with the dc electrode the oscillating voltage is applied to and therefore experiences the largest electric field. At a distance of 450\,\textmu m, $\Omega_\textrm{eff}$ is reduced by a factor of 7. The blue dashed line indicates the expected behaviour based on a simulation of a voltage applied to a single electrode. The error bars are smaller than the data points.}
\end{figure}

Lastly, we investigate the localisation of the electric field. We apply the oscillating voltage to a fixed electrode and measure $\Omega_\textrm{eff}$ at different positions along the trap axis as shown in Fig.~\ref{fig_e_field}. At each point, we ensure that the mode orientations and frequencies are kept constant by adjusting the static confining potential.  Figure~\ref{fig_e_field} also includes simulation data based on a trap model that only considers a voltage applied to a single electrode. The discrepancy between the data and the model is consistent with pickup on nearby electrodes.  We expect that this pickup can be significantly reduced by careful design of the trap structures and the electrical routing path to minimise capacitive couplings between electrodes. About 450\,\textmu m away from its initial position, we observe a factor of 7 suppression in $\Omega_\textrm{eff}$. This suppression can be increased by moving the ion further away, or driving two neighbouring dc electrodes simultaneously out of phase, which would reduce the electric field projection onto the radial modes of ions in distant zones. The orientation and frequency of the motional modes in each zone also provide additional degrees of freedom to reduce the strength of the interaction in the non-target zone. Additionally, this crosstalk results in a coherent error and can be corrected with additional pulses.

In this proof-of-principle demonstration, $\Omega_\textrm{eff}$ is limited to $\sim10\,$kHz partly due to the low amplitude of the applied gradient. For example, increasing the microwave gradient to $100$\,T/m would increase $\Omega_\textrm{eff}$ to $\sim300\,$kHz with our parameters. Such large gradients are already a requirement for fast, high-fidelity two-qubit gates. Increasing the amplitude of the forced ion motion can also increase $\Omega_\textrm{eff}$. One can apply a larger oscillating voltage to the driving electrode or tune $\omega_e$ closer to a motional mode frequency. The larger ion motion increases the sensitivity to anharmonicities in the trap potential, which can lead to additional errors. Tuning closer to resonance increases the pulse shaping requirements~\cite{supplementary} and the sensitivity to mode frequency fluctuations. 

Thus far, we have only discussed the use of this method for addressing qubits in separate zones. To address qubits within the same zone, an electric field driving modes other than the centre-of-mass mode could be used. For example, a differential electric field that drives the forced motion sideband on the out-of-phase mode of a two-ion crystal can be used to generate a differential Rabi frequency between the two ions.

Finally, while the effective Hamiltonian in Eq.~\ref{eq_sf} acts only on the qubit, ion temperature has a second-order influence on the dynamics~\cite{Sutherland2022}. We verify numerically that this effect results in infidelities below the $10^{-4}$ level for an ion temperature of $\bar{n}=1$. Our estimate is limited by the number of Fock states included in the simulation~\cite{supplementary}. Further numerical and experimental work is necessary to verify the performance of the forced motion sideband, especially for ions in anharmonic potentials. However, such errors can be mitigated by reducing the magnitude of the electric field or its detuning at the cost of the strength of the interaction. As reducing the motional excitation of the ion decreases this effect and the second-order temperature dependence, we expect that these errors can be reduced to an arbitrary degree by decreasing the magnitude of the electric field or its detuning at the cost of the strength of the interaction.

In conclusion, we have introduced a new method of driving single-qubit rotations in trapped ions using a spin-dependent gradient and an electric field. The electric field can be used to control both the amplitude and phase of the single-qubit rotations. We have demonstrated this method experimentally, generating the spin-dependent gradient with a microwave field from an integrated current-carrying electrode or a laser on a quadrupole transition. This method is also applicable to two-photon Raman transitions~\cite{Wineland1998} and can be used with other charged particles such as molecular~\cite{Chou2017} or highly charged ions~\cite{Kozlov2018, Micke2020}, or electrons~\cite{Matthiesen2021}.  With the local control offered by the electric field, a single gradient could drive single-qubit rotations simultaneously in multiple zones of a trapped-ion quantum processor, each with their own controllable amplitude, phase, and even frequency. The use of a single gradient is readily applicable to experimental implementations that use a single laser source routed into different zones for parallel operations~\cite{Leibfried2007}, or a global magnetic-field gradient from current-carrying electrodes.  Further, this scheme imposes minimal additional hardware requirements and leverages the existing infrastructure in modern ion traps, namely applying fast voltage waveforms to dc electrodes for shuttling~\cite{Bowler2012, Pino2020}. Future work will focus on increasing the speed of these operations and characterising parallel operations on ions in separate zones.

We would like to thank David Wineland for insightful discussions, and are grateful to the entire Oxford Ionics team for their support.

\bibliographystyle{apsrev4-1}
\bibliography{refs}

\begin{thebibliography}{36}%
\makeatletter
\providecommand \@ifxundefined [1]{%
 \@ifx{#1\undefined}
}%
\providecommand \@ifnum [1]{%
 \ifnum #1\expandafter \@firstoftwo
 \else \expandafter \@secondoftwo
 \fi
}%
\providecommand \@ifx [1]{%
 \ifx #1\expandafter \@firstoftwo
 \else \expandafter \@secondoftwo
 \fi
}%
\providecommand \natexlab [1]{#1}%
\providecommand \enquote  [1]{``#1''}%
\providecommand \bibnamefont  [1]{#1}%
\providecommand \bibfnamefont [1]{#1}%
\providecommand \citenamefont [1]{#1}%
\providecommand \href@noop [0]{\@secondoftwo}%
\providecommand \href [0]{\begingroup \@sanitize@url \@href}%
\providecommand \@href[1]{\@@startlink{#1}\@@href}%
\providecommand \@@href[1]{\endgroup#1\@@endlink}%
\providecommand \@sanitize@url [0]{\catcode `\\12\catcode `\$12\catcode
  `\&12\catcode `\#12\catcode `\^12\catcode `\_12\catcode `\%12\relax}%
\providecommand \@@startlink[1]{}%
\providecommand \@@endlink[0]{}%
\providecommand \url  [0]{\begingroup\@sanitize@url \@url }%
\providecommand \@url [1]{\endgroup\@href {#1}{\urlprefix }}%
\providecommand \urlprefix  [0]{URL }%
\providecommand \Eprint [0]{\href }%
\providecommand \doibase [0]{http://dx.doi.org/}%
\providecommand \selectlanguage [0]{\@gobble}%
\providecommand \bibinfo  [0]{\@secondoftwo}%
\providecommand \bibfield  [0]{\@secondoftwo}%
\providecommand \translation [1]{[#1]}%
\providecommand \BibitemOpen [0]{}%
\providecommand \bibitemStop [0]{}%
\providecommand \bibitemNoStop [0]{.\EOS\space}%
\providecommand \EOS [0]{\spacefactor3000\relax}%
\providecommand \BibitemShut  [1]{\csname bibitem#1\endcsname}%
\let\auto@bib@innerbib\@empty
\bibitem [{\citenamefont {Langer}\ \emph {et~al.}(2005)\citenamefont {Langer},
  \citenamefont {Ozeri}, \citenamefont {Jost}, \citenamefont {Chiaverini},
  \citenamefont {DeMarco}, \citenamefont {Ben-Kish}, \citenamefont {Blakestad},
  \citenamefont {Britton}, \citenamefont {Hume}, \citenamefont {Itano},
  \citenamefont {Leibfried}, \citenamefont {Reichle}, \citenamefont
  {Rosenband}, \citenamefont {Schaetz}, \citenamefont {Schmidt},\ and\
  \citenamefont {Wineland}}]{Langer2005}%
  \BibitemOpen
  \bibfield  {author} {\bibinfo {author} {\bibfnamefont {C.}~\bibnamefont
  {Langer}}, \bibinfo {author} {\bibfnamefont {R.}~\bibnamefont {Ozeri}},
  \bibinfo {author} {\bibfnamefont {J.~D.}\ \bibnamefont {Jost}}, \bibinfo
  {author} {\bibfnamefont {J.}~\bibnamefont {Chiaverini}}, \bibinfo {author}
  {\bibfnamefont {B.}~\bibnamefont {DeMarco}}, \bibinfo {author} {\bibfnamefont
  {A.}~\bibnamefont {Ben-Kish}}, \bibinfo {author} {\bibfnamefont {R.~B.}\
  \bibnamefont {Blakestad}}, \bibinfo {author} {\bibfnamefont {J.}~\bibnamefont
  {Britton}}, \bibinfo {author} {\bibfnamefont {D.~B.}\ \bibnamefont {Hume}},
  \bibinfo {author} {\bibfnamefont {W.~M.}\ \bibnamefont {Itano}}, \bibinfo
  {author} {\bibfnamefont {D.}~\bibnamefont {Leibfried}}, \bibinfo {author}
  {\bibfnamefont {R.}~\bibnamefont {Reichle}}, \bibinfo {author} {\bibfnamefont
  {T.}~\bibnamefont {Rosenband}}, \bibinfo {author} {\bibfnamefont
  {T.}~\bibnamefont {Schaetz}}, \bibinfo {author} {\bibfnamefont {P.~O.}\
  \bibnamefont {Schmidt}}, \ and\ \bibinfo {author} {\bibfnamefont {D.~J.}\
  \bibnamefont {Wineland}},\ }\href {\doibase 10.1103/PhysRevLett.95.060502}
  {\bibfield  {journal} {\bibinfo  {journal} {Phys. Rev. Lett.}\ }\textbf
  {\bibinfo {volume} {95}},\ \bibinfo {pages} {060502} (\bibinfo {year}
  {2005})}\BibitemShut {NoStop}%
\bibitem [{\citenamefont {Wang}\ \emph {et~al.}(2021)\citenamefont {Wang},
  \citenamefont {Luan}, \citenamefont {Qiao}, \citenamefont {Um}, \citenamefont
  {Zhang}, \citenamefont {Wang}, \citenamefont {Yuan}, \citenamefont {Gu},
  \citenamefont {Zhang},\ and\ \citenamefont {Kim}}]{Wang2021}%
  \BibitemOpen
  \bibfield  {author} {\bibinfo {author} {\bibfnamefont {P.}~\bibnamefont
  {Wang}}, \bibinfo {author} {\bibfnamefont {C.-Y.}\ \bibnamefont {Luan}},
  \bibinfo {author} {\bibfnamefont {M.}~\bibnamefont {Qiao}}, \bibinfo {author}
  {\bibfnamefont {M.}~\bibnamefont {Um}}, \bibinfo {author} {\bibfnamefont
  {J.}~\bibnamefont {Zhang}}, \bibinfo {author} {\bibfnamefont
  {Y.}~\bibnamefont {Wang}}, \bibinfo {author} {\bibfnamefont {X.}~\bibnamefont
  {Yuan}}, \bibinfo {author} {\bibfnamefont {M.}~\bibnamefont {Gu}}, \bibinfo
  {author} {\bibfnamefont {J.}~\bibnamefont {Zhang}}, \ and\ \bibinfo {author}
  {\bibfnamefont {K.}~\bibnamefont {Kim}},\ }\href@noop {} {\bibfield
  {journal} {\bibinfo  {journal} {Nat. Commun.}\ }\textbf {\bibinfo {volume}
  {12}},\ \bibinfo {pages} {233} (\bibinfo {year} {2021})}\BibitemShut
  {NoStop}%
\bibitem [{\citenamefont {Harty}\ \emph {et~al.}(2014)\citenamefont {Harty},
  \citenamefont {Allcock}, \citenamefont {Ballance}, \citenamefont {Guidoni},
  \citenamefont {Janacek}, \citenamefont {Linke}, \citenamefont {Stacey},\ and\
  \citenamefont {Lucas}}]{Harty2014}%
  \BibitemOpen
  \bibfield  {author} {\bibinfo {author} {\bibfnamefont {T.~P.}\ \bibnamefont
  {Harty}}, \bibinfo {author} {\bibfnamefont {D.~T.~C.}\ \bibnamefont
  {Allcock}}, \bibinfo {author} {\bibfnamefont {C.~J.}\ \bibnamefont
  {Ballance}}, \bibinfo {author} {\bibfnamefont {L.}~\bibnamefont {Guidoni}},
  \bibinfo {author} {\bibfnamefont {H.~A.}\ \bibnamefont {Janacek}}, \bibinfo
  {author} {\bibfnamefont {N.~M.}\ \bibnamefont {Linke}}, \bibinfo {author}
  {\bibfnamefont {D.~N.}\ \bibnamefont {Stacey}}, \ and\ \bibinfo {author}
  {\bibfnamefont {D.~M.}\ \bibnamefont {Lucas}},\ }\href {\doibase
  10.1103/PhysRevLett.113.220501} {\bibfield  {journal} {\bibinfo  {journal}
  {Phys. Rev. Lett.}\ }\textbf {\bibinfo {volume} {113}},\ \bibinfo {pages}
  {220501} (\bibinfo {year} {2014})}\BibitemShut {NoStop}%
\bibitem [{\citenamefont {Ballance}\ \emph {et~al.}(2016)\citenamefont
  {Ballance}, \citenamefont {Harty}, \citenamefont {Linke}, \citenamefont
  {Sepiol},\ and\ \citenamefont {Lucas}}]{Ballance2016}%
  \BibitemOpen
  \bibfield  {author} {\bibinfo {author} {\bibfnamefont {C.~J.}\ \bibnamefont
  {Ballance}}, \bibinfo {author} {\bibfnamefont {T.~P.}\ \bibnamefont {Harty}},
  \bibinfo {author} {\bibfnamefont {N.~M.}\ \bibnamefont {Linke}}, \bibinfo
  {author} {\bibfnamefont {M.~A.}\ \bibnamefont {Sepiol}}, \ and\ \bibinfo
  {author} {\bibfnamefont {D.~M.}\ \bibnamefont {Lucas}},\ }\href {\doibase
  10.1103/PhysRevLett.117.060504} {\bibfield  {journal} {\bibinfo  {journal}
  {Phys. Rev. Lett.}\ }\textbf {\bibinfo {volume} {117}},\ \bibinfo {pages}
  {060504} (\bibinfo {year} {2016})}\BibitemShut {NoStop}%
\bibitem [{\citenamefont {Gaebler}\ \emph {et~al.}(2016)\citenamefont
  {Gaebler}, \citenamefont {Tan}, \citenamefont {Lin}, \citenamefont {Wan},
  \citenamefont {Bowler}, \citenamefont {Keith}, \citenamefont {Glancy},
  \citenamefont {Coakley}, \citenamefont {Knill}, \citenamefont {Leibfried},\
  and\ \citenamefont {Wineland}}]{Gaebler2016}%
  \BibitemOpen
  \bibfield  {author} {\bibinfo {author} {\bibfnamefont {J.~P.}\ \bibnamefont
  {Gaebler}}, \bibinfo {author} {\bibfnamefont {T.~R.}\ \bibnamefont {Tan}},
  \bibinfo {author} {\bibfnamefont {Y.}~\bibnamefont {Lin}}, \bibinfo {author}
  {\bibfnamefont {Y.}~\bibnamefont {Wan}}, \bibinfo {author} {\bibfnamefont
  {R.}~\bibnamefont {Bowler}}, \bibinfo {author} {\bibfnamefont {A.~C.}\
  \bibnamefont {Keith}}, \bibinfo {author} {\bibfnamefont {S.}~\bibnamefont
  {Glancy}}, \bibinfo {author} {\bibfnamefont {K.}~\bibnamefont {Coakley}},
  \bibinfo {author} {\bibfnamefont {E.}~\bibnamefont {Knill}}, \bibinfo
  {author} {\bibfnamefont {D.}~\bibnamefont {Leibfried}}, \ and\ \bibinfo
  {author} {\bibfnamefont {D.~J.}\ \bibnamefont {Wineland}},\ }\href {\doibase
  10.1103/PhysRevLett.117.060505} {\bibfield  {journal} {\bibinfo  {journal}
  {Phys. Rev. Lett.}\ }\textbf {\bibinfo {volume} {117}},\ \bibinfo {pages}
  {060505} (\bibinfo {year} {2016})}\BibitemShut {NoStop}%
\bibitem [{\citenamefont {Srinivas}\ \emph {et~al.}(2021)\citenamefont
  {Srinivas}, \citenamefont {Burd}, \citenamefont {Knaack}, \citenamefont
  {Sutherland}, \citenamefont {Kwiatkowski}, \citenamefont {Glancy},
  \citenamefont {Knill}, \citenamefont {Wineland}, \citenamefont {Leibfried},
  \citenamefont {Wilson}, \citenamefont {Allcock},\ and\ \citenamefont
  {Slichter}}]{Srinivas2021}%
  \BibitemOpen
  \bibfield  {author} {\bibinfo {author} {\bibfnamefont {R.}~\bibnamefont
  {Srinivas}}, \bibinfo {author} {\bibfnamefont {S.~C.}\ \bibnamefont {Burd}},
  \bibinfo {author} {\bibfnamefont {H.~M.}\ \bibnamefont {Knaack}}, \bibinfo
  {author} {\bibfnamefont {R.~T.}\ \bibnamefont {Sutherland}}, \bibinfo
  {author} {\bibfnamefont {A.}~\bibnamefont {Kwiatkowski}}, \bibinfo {author}
  {\bibfnamefont {S.}~\bibnamefont {Glancy}}, \bibinfo {author} {\bibfnamefont
  {E.}~\bibnamefont {Knill}}, \bibinfo {author} {\bibfnamefont {D.~J.}\
  \bibnamefont {Wineland}}, \bibinfo {author} {\bibfnamefont {D.}~\bibnamefont
  {Leibfried}}, \bibinfo {author} {\bibfnamefont {A.~C.}\ \bibnamefont
  {Wilson}}, \bibinfo {author} {\bibfnamefont {D.~T.~C.}\ \bibnamefont
  {Allcock}}, \ and\ \bibinfo {author} {\bibfnamefont {D.~H.}\ \bibnamefont
  {Slichter}},\ }\href@noop {} {\bibfield  {journal} {\bibinfo  {journal}
  {Nature}\ }\textbf {\bibinfo {volume} {597}},\ \bibinfo {pages} {209}
  (\bibinfo {year} {2021})}\BibitemShut {NoStop}%
\bibitem [{\citenamefont {Clark}\ \emph {et~al.}(2021)\citenamefont {Clark},
  \citenamefont {Tinkey}, \citenamefont {Sawyer}, \citenamefont {Meier},
  \citenamefont {Burkhardt}, \citenamefont {Seck}, \citenamefont {Shappert},
  \citenamefont {Guise}, \citenamefont {Volin}, \citenamefont {Fallek},
  \citenamefont {Hayden}, \citenamefont {Rellergert},\ and\ \citenamefont
  {Brown}}]{Clark2021}%
  \BibitemOpen
  \bibfield  {author} {\bibinfo {author} {\bibfnamefont {C.~R.}\ \bibnamefont
  {Clark}}, \bibinfo {author} {\bibfnamefont {H.~N.}\ \bibnamefont {Tinkey}},
  \bibinfo {author} {\bibfnamefont {B.~C.}\ \bibnamefont {Sawyer}}, \bibinfo
  {author} {\bibfnamefont {A.~M.}\ \bibnamefont {Meier}}, \bibinfo {author}
  {\bibfnamefont {K.~A.}\ \bibnamefont {Burkhardt}}, \bibinfo {author}
  {\bibfnamefont {C.~M.}\ \bibnamefont {Seck}}, \bibinfo {author}
  {\bibfnamefont {C.~M.}\ \bibnamefont {Shappert}}, \bibinfo {author}
  {\bibfnamefont {N.~D.}\ \bibnamefont {Guise}}, \bibinfo {author}
  {\bibfnamefont {C.~E.}\ \bibnamefont {Volin}}, \bibinfo {author}
  {\bibfnamefont {S.~D.}\ \bibnamefont {Fallek}}, \bibinfo {author}
  {\bibfnamefont {H.~T.}\ \bibnamefont {Hayden}}, \bibinfo {author}
  {\bibfnamefont {W.~G.}\ \bibnamefont {Rellergert}}, \ and\ \bibinfo {author}
  {\bibfnamefont {K.~R.}\ \bibnamefont {Brown}},\ }\href {\doibase
  10.1103/PhysRevLett.127.130505} {\bibfield  {journal} {\bibinfo  {journal}
  {Phys. Rev. Lett.}\ }\textbf {\bibinfo {volume} {127}},\ \bibinfo {pages}
  {130505} (\bibinfo {year} {2021})}\BibitemShut {NoStop}%
\bibitem [{\citenamefont {Wineland}\ \emph {et~al.}(1998)\citenamefont
  {Wineland}, \citenamefont {Monroe}, \citenamefont {Itano}, \citenamefont
  {Leibfried}, \citenamefont {King},\ and\ \citenamefont
  {Meekhof}}]{Wineland1998}%
  \BibitemOpen
  \bibfield  {author} {\bibinfo {author} {\bibfnamefont {D.~J.}\ \bibnamefont
  {Wineland}}, \bibinfo {author} {\bibfnamefont {C.}~\bibnamefont {Monroe}},
  \bibinfo {author} {\bibfnamefont {W.~M.}\ \bibnamefont {Itano}}, \bibinfo
  {author} {\bibfnamefont {D.}~\bibnamefont {Leibfried}}, \bibinfo {author}
  {\bibfnamefont {B.~E.}\ \bibnamefont {King}}, \ and\ \bibinfo {author}
  {\bibfnamefont {D.~M.}\ \bibnamefont {Meekhof}},\ }\href {\doibase
  10.6028/jres.103.019} {\bibfield  {journal} {\bibinfo  {journal} {J. Res.
  Natl. Inst. Stand. Technol.}\ }\textbf {\bibinfo {volume} {103}},\ \bibinfo
  {pages} {259} (\bibinfo {year} {1998})}\BibitemShut {NoStop}%
\bibitem [{\citenamefont {Kielpinski}\ \emph {et~al.}(2002)\citenamefont
  {Kielpinski}, \citenamefont {Monroe},\ and\ \citenamefont
  {Wineland}}]{Kielpinski2002}%
  \BibitemOpen
  \bibfield  {author} {\bibinfo {author} {\bibfnamefont {D.}~\bibnamefont
  {Kielpinski}}, \bibinfo {author} {\bibfnamefont {C.}~\bibnamefont {Monroe}},
  \ and\ \bibinfo {author} {\bibfnamefont {D.~J.}\ \bibnamefont {Wineland}},\
  }\href {\doibase 10.1038/nature00784} {\bibfield  {journal} {\bibinfo
  {journal} {Nature}\ }\textbf {\bibinfo {volume} {417}},\ \bibinfo {pages}
  {709} (\bibinfo {year} {2002})}\BibitemShut {NoStop}%
\bibitem [{\citenamefont {Barenco}\ \emph {et~al.}(1995)\citenamefont
  {Barenco}, \citenamefont {Bennett}, \citenamefont {Cleve}, \citenamefont
  {DiVincenzo}, \citenamefont {Margolus}, \citenamefont {Shor}, \citenamefont
  {Sleator}, \citenamefont {Smolin},\ and\ \citenamefont
  {Weinfurter}}]{Barenco1995}%
  \BibitemOpen
  \bibfield  {author} {\bibinfo {author} {\bibfnamefont {A.}~\bibnamefont
  {Barenco}}, \bibinfo {author} {\bibfnamefont {C.~H.}\ \bibnamefont
  {Bennett}}, \bibinfo {author} {\bibfnamefont {R.}~\bibnamefont {Cleve}},
  \bibinfo {author} {\bibfnamefont {D.~P.}\ \bibnamefont {DiVincenzo}},
  \bibinfo {author} {\bibfnamefont {N.}~\bibnamefont {Margolus}}, \bibinfo
  {author} {\bibfnamefont {P.}~\bibnamefont {Shor}}, \bibinfo {author}
  {\bibfnamefont {T.}~\bibnamefont {Sleator}}, \bibinfo {author} {\bibfnamefont
  {J.~A.}\ \bibnamefont {Smolin}}, \ and\ \bibinfo {author} {\bibfnamefont
  {H.}~\bibnamefont {Weinfurter}},\ }\href {\doibase 10.1103/PhysRevA.52.3457}
  {\bibfield  {journal} {\bibinfo  {journal} {Phys. Rev. A}\ }\textbf {\bibinfo
  {volume} {52}},\ \bibinfo {pages} {3457} (\bibinfo {year}
  {1995})}\BibitemShut {NoStop}%
\bibitem [{\citenamefont {Aharonov}\ and\ \citenamefont
  {Ben-Or}(1996)}]{Aharonov1996}%
  \BibitemOpen
  \bibfield  {author} {\bibinfo {author} {\bibfnamefont {D.}~\bibnamefont
  {Aharonov}}\ and\ \bibinfo {author} {\bibfnamefont {M.}~\bibnamefont
  {Ben-Or}},\ }in\ \href@noop {} {\emph {\bibinfo {booktitle} {Proceedings of
  37th Conference on Foundations of Computer Science}}}\ (\bibinfo
  {organization} {IEEE},\ \bibinfo {year} {1996})\ pp.\ \bibinfo {pages}
  {46--55}\BibitemShut {NoStop}%
\bibitem [{\citenamefont {N\"agerl}\ \emph {et~al.}(1999)\citenamefont
  {N\"agerl}, \citenamefont {Leibfried}, \citenamefont {Rohde}, \citenamefont
  {Thalhammer}, \citenamefont {Eschner}, \citenamefont {Schmidt-Kaler},\ and\
  \citenamefont {Blatt}}]{Naegerl1999}%
  \BibitemOpen
  \bibfield  {author} {\bibinfo {author} {\bibfnamefont {H.~C.}\ \bibnamefont
  {N\"agerl}}, \bibinfo {author} {\bibfnamefont {D.}~\bibnamefont {Leibfried}},
  \bibinfo {author} {\bibfnamefont {H.}~\bibnamefont {Rohde}}, \bibinfo
  {author} {\bibfnamefont {G.}~\bibnamefont {Thalhammer}}, \bibinfo {author}
  {\bibfnamefont {J.}~\bibnamefont {Eschner}}, \bibinfo {author} {\bibfnamefont
  {F.}~\bibnamefont {Schmidt-Kaler}}, \ and\ \bibinfo {author} {\bibfnamefont
  {R.}~\bibnamefont {Blatt}},\ }\href@noop {} {\bibfield  {journal} {\bibinfo
  {journal} {Phys. Rev. A}\ }\textbf {\bibinfo {volume} {60}},\ \bibinfo
  {pages} {145} (\bibinfo {year} {1999})}\BibitemShut {NoStop}%
\bibitem [{\citenamefont {Niffenegger}\ \emph {et~al.}(2020)\citenamefont
  {Niffenegger}, \citenamefont {Stuart}, \citenamefont {Sorace-Agaskar},
  \citenamefont {Kharas}, \citenamefont {Bramhavar}, \citenamefont {Bruzewicz},
  \citenamefont {Loh}, \citenamefont {Maxson}, \citenamefont {McConnell},
  \citenamefont {Reens}, \citenamefont {West}, \citenamefont {Sage},\ and\
  \citenamefont {Chiaverini}}]{Niffenegger2020}%
  \BibitemOpen
  \bibfield  {author} {\bibinfo {author} {\bibfnamefont {R.~J.}\ \bibnamefont
  {Niffenegger}}, \bibinfo {author} {\bibfnamefont {J.}~\bibnamefont {Stuart}},
  \bibinfo {author} {\bibfnamefont {C.}~\bibnamefont {Sorace-Agaskar}},
  \bibinfo {author} {\bibfnamefont {D.}~\bibnamefont {Kharas}}, \bibinfo
  {author} {\bibfnamefont {S.}~\bibnamefont {Bramhavar}}, \bibinfo {author}
  {\bibfnamefont {C.~D.}\ \bibnamefont {Bruzewicz}}, \bibinfo {author}
  {\bibfnamefont {W.}~\bibnamefont {Loh}}, \bibinfo {author} {\bibfnamefont
  {R.~T.}\ \bibnamefont {Maxson}}, \bibinfo {author} {\bibfnamefont
  {R.}~\bibnamefont {McConnell}}, \bibinfo {author} {\bibfnamefont
  {D.}~\bibnamefont {Reens}}, \bibinfo {author} {\bibfnamefont {G.~N.}\
  \bibnamefont {West}}, \bibinfo {author} {\bibfnamefont {J.~M.}\ \bibnamefont
  {Sage}}, \ and\ \bibinfo {author} {\bibfnamefont {J.}~\bibnamefont
  {Chiaverini}},\ }\href@noop {} {\bibfield  {journal} {\bibinfo  {journal}
  {Nature}\ }\textbf {\bibinfo {volume} {586}},\ \bibinfo {pages} {538}
  (\bibinfo {year} {2020})}\BibitemShut {NoStop}%
\bibitem [{\citenamefont {Mehta}\ \emph {et~al.}(2020)\citenamefont {Mehta},
  \citenamefont {Zhang}, \citenamefont {Malinowski}, \citenamefont {Nguyen},
  \citenamefont {Stadler},\ and\ \citenamefont {Home}}]{Mehta2020}%
  \BibitemOpen
  \bibfield  {author} {\bibinfo {author} {\bibfnamefont {K.~K.}\ \bibnamefont
  {Mehta}}, \bibinfo {author} {\bibfnamefont {C.}~\bibnamefont {Zhang}},
  \bibinfo {author} {\bibfnamefont {M.}~\bibnamefont {Malinowski}}, \bibinfo
  {author} {\bibfnamefont {T.-L.}\ \bibnamefont {Nguyen}}, \bibinfo {author}
  {\bibfnamefont {M.}~\bibnamefont {Stadler}}, \ and\ \bibinfo {author}
  {\bibfnamefont {J.~P.}\ \bibnamefont {Home}},\ }\href@noop {} {\bibfield
  {journal} {\bibinfo  {journal} {Nature}\ }\textbf {\bibinfo {volume} {586}},\
  \bibinfo {pages} {533} (\bibinfo {year} {2020})}\BibitemShut {NoStop}%
\bibitem [{\citenamefont {Brown}\ \emph {et~al.}(2011)\citenamefont {Brown},
  \citenamefont {Wilson}, \citenamefont {Colombe}, \citenamefont {Ospelkaus},
  \citenamefont {Meier}, \citenamefont {Knill}, \citenamefont {Leibfried},\
  and\ \citenamefont {Wineland}}]{Brown2011}%
  \BibitemOpen
  \bibfield  {author} {\bibinfo {author} {\bibfnamefont {K.~R.}\ \bibnamefont
  {Brown}}, \bibinfo {author} {\bibfnamefont {A.~C.}\ \bibnamefont {Wilson}},
  \bibinfo {author} {\bibfnamefont {Y.}~\bibnamefont {Colombe}}, \bibinfo
  {author} {\bibfnamefont {C.}~\bibnamefont {Ospelkaus}}, \bibinfo {author}
  {\bibfnamefont {A.~M.}\ \bibnamefont {Meier}}, \bibinfo {author}
  {\bibfnamefont {E.}~\bibnamefont {Knill}}, \bibinfo {author} {\bibfnamefont
  {D.}~\bibnamefont {Leibfried}}, \ and\ \bibinfo {author} {\bibfnamefont
  {D.~J.}\ \bibnamefont {Wineland}},\ }\href {\doibase
  doi.org/10.1103/PhysRevA.84.030303} {\bibfield  {journal} {\bibinfo
  {journal} {Phys. Rev. A}\ }\textbf {\bibinfo {volume} {84}},\ \bibinfo
  {pages} {030303} (\bibinfo {year} {2011})}\BibitemShut {NoStop}%
\bibitem [{\citenamefont {Ospelkaus}\ \emph {et~al.}(2008)\citenamefont
  {Ospelkaus}, \citenamefont {Langer}, \citenamefont {Amini}, \citenamefont
  {Brown}, \citenamefont {Leibfried},\ and\ \citenamefont
  {Wineland}}]{Ospelkaus2008}%
  \BibitemOpen
  \bibfield  {author} {\bibinfo {author} {\bibfnamefont {C.}~\bibnamefont
  {Ospelkaus}}, \bibinfo {author} {\bibfnamefont {C.~E.}\ \bibnamefont
  {Langer}}, \bibinfo {author} {\bibfnamefont {J.~M.}\ \bibnamefont {Amini}},
  \bibinfo {author} {\bibfnamefont {K.~R.}\ \bibnamefont {Brown}}, \bibinfo
  {author} {\bibfnamefont {D.}~\bibnamefont {Leibfried}}, \ and\ \bibinfo
  {author} {\bibfnamefont {D.~J.}\ \bibnamefont {Wineland}},\ }\href {\doibase
  10.1103/PhysRevLett.101.090502} {\bibfield  {journal} {\bibinfo  {journal}
  {Phys. Rev. Lett.}\ }\textbf {\bibinfo {volume} {101}},\ \bibinfo {pages}
  {090502} (\bibinfo {year} {2008})}\BibitemShut {NoStop}%
\bibitem [{\citenamefont {Mintert}\ and\ \citenamefont
  {Wunderlich}(2001)}]{Mintert2001}%
  \BibitemOpen
  \bibfield  {author} {\bibinfo {author} {\bibfnamefont {F.}~\bibnamefont
  {Mintert}}\ and\ \bibinfo {author} {\bibfnamefont {C.}~\bibnamefont
  {Wunderlich}},\ }\href {\doibase 10.1103/PhysRevLett.87.257904} {\bibfield
  {journal} {\bibinfo  {journal} {Phys. Rev. Lett.}\ }\textbf {\bibinfo
  {volume} {87}},\ \bibinfo {pages} {257904} (\bibinfo {year}
  {2001})}\BibitemShut {NoStop}%
\bibitem [{\citenamefont {Piltz}\ \emph {et~al.}(2014)\citenamefont {Piltz},
  \citenamefont {Sriarunothai}, \citenamefont {Var{\'o}n},\ and\ \citenamefont
  {Wunderlich}}]{Piltz2014}%
  \BibitemOpen
  \bibfield  {author} {\bibinfo {author} {\bibfnamefont {C.}~\bibnamefont
  {Piltz}}, \bibinfo {author} {\bibfnamefont {T.}~\bibnamefont {Sriarunothai}},
  \bibinfo {author} {\bibfnamefont {A.~F.}\ \bibnamefont {Var{\'o}n}}, \ and\
  \bibinfo {author} {\bibfnamefont {C.}~\bibnamefont {Wunderlich}},\
  }\href@noop {} {\bibfield  {journal} {\bibinfo  {journal} {Nat. Commun.}\
  }\textbf {\bibinfo {volume} {5}},\ \bibinfo {pages} {1} (\bibinfo {year}
  {2014})}\BibitemShut {NoStop}%
\bibitem [{\citenamefont {Warring}\ \emph
  {et~al.}(2013{\natexlab{a}})\citenamefont {Warring}, \citenamefont
  {Ospelkaus}, \citenamefont {Colombe}, \citenamefont {Brown}, \citenamefont
  {Amini}, \citenamefont {Carsjens}, \citenamefont {Leibfried},\ and\
  \citenamefont {Wineland}}]{Warring2013b}%
  \BibitemOpen
  \bibfield  {author} {\bibinfo {author} {\bibfnamefont {U.}~\bibnamefont
  {Warring}}, \bibinfo {author} {\bibfnamefont {C.}~\bibnamefont {Ospelkaus}},
  \bibinfo {author} {\bibfnamefont {Y.}~\bibnamefont {Colombe}}, \bibinfo
  {author} {\bibfnamefont {K.~R.}\ \bibnamefont {Brown}}, \bibinfo {author}
  {\bibfnamefont {J.~M.}\ \bibnamefont {Amini}}, \bibinfo {author}
  {\bibfnamefont {M.}~\bibnamefont {Carsjens}}, \bibinfo {author}
  {\bibfnamefont {D.}~\bibnamefont {Leibfried}}, \ and\ \bibinfo {author}
  {\bibfnamefont {D.~J.}\ \bibnamefont {Wineland}},\ }\href {\doibase
  10.1103/PhysRevA.87.013437} {\bibfield  {journal} {\bibinfo  {journal} {Phys.
  Rev. A}\ }\textbf {\bibinfo {volume} {87}},\ \bibinfo {pages} {013437}
  (\bibinfo {year} {2013}{\natexlab{a}})}\BibitemShut {NoStop}%
\bibitem [{\citenamefont {Aude~Craik}\ \emph {et~al.}(2017)\citenamefont
  {Aude~Craik}, \citenamefont {Linke}, \citenamefont {Sepiol}, \citenamefont
  {Harty}, \citenamefont {Goodwin}, \citenamefont {Ballance}, \citenamefont
  {Stacey}, \citenamefont {Steane}, \citenamefont {Lucas},\ and\ \citenamefont
  {Allcock}}]{Craik2017}%
  \BibitemOpen
  \bibfield  {author} {\bibinfo {author} {\bibfnamefont {D.~P.~L.}\
  \bibnamefont {Aude~Craik}}, \bibinfo {author} {\bibfnamefont {N.~M.}\
  \bibnamefont {Linke}}, \bibinfo {author} {\bibfnamefont {M.~A.}\ \bibnamefont
  {Sepiol}}, \bibinfo {author} {\bibfnamefont {T.~P.}\ \bibnamefont {Harty}},
  \bibinfo {author} {\bibfnamefont {J.~F.}\ \bibnamefont {Goodwin}}, \bibinfo
  {author} {\bibfnamefont {C.~J.}\ \bibnamefont {Ballance}}, \bibinfo {author}
  {\bibfnamefont {D.~N.}\ \bibnamefont {Stacey}}, \bibinfo {author}
  {\bibfnamefont {A.~M.}\ \bibnamefont {Steane}}, \bibinfo {author}
  {\bibfnamefont {D.~M.}\ \bibnamefont {Lucas}}, \ and\ \bibinfo {author}
  {\bibfnamefont {D.~T.~C.}\ \bibnamefont {Allcock}},\ }\href@noop {}
  {\bibfield  {journal} {\bibinfo  {journal} {Phys. Rev. A}\ }\textbf {\bibinfo
  {volume} {95}},\ \bibinfo {pages} {022337} (\bibinfo {year}
  {2017})}\BibitemShut {NoStop}%
\bibitem [{\citenamefont {Dong}\ \emph {et~al.}(2022)\citenamefont {Dong},
  \citenamefont {Clark}, \citenamefont {Leenheer}, \citenamefont {Zimmermann},
  \citenamefont {Dominguez}, \citenamefont {Menssen}, \citenamefont {Heim},
  \citenamefont {Gilbert}, \citenamefont {Englund},\ and\ \citenamefont
  {Eichenfield}}]{Dong2022}%
  \BibitemOpen
  \bibfield  {author} {\bibinfo {author} {\bibfnamefont {M.}~\bibnamefont
  {Dong}}, \bibinfo {author} {\bibfnamefont {G.}~\bibnamefont {Clark}},
  \bibinfo {author} {\bibfnamefont {A.~J.}\ \bibnamefont {Leenheer}}, \bibinfo
  {author} {\bibfnamefont {M.}~\bibnamefont {Zimmermann}}, \bibinfo {author}
  {\bibfnamefont {D.}~\bibnamefont {Dominguez}}, \bibinfo {author}
  {\bibfnamefont {A.~J.}\ \bibnamefont {Menssen}}, \bibinfo {author}
  {\bibfnamefont {D.}~\bibnamefont {Heim}}, \bibinfo {author} {\bibfnamefont
  {G.}~\bibnamefont {Gilbert}}, \bibinfo {author} {\bibfnamefont
  {D.}~\bibnamefont {Englund}}, \ and\ \bibinfo {author} {\bibfnamefont
  {M.}~\bibnamefont {Eichenfield}},\ }\href@noop {} {\bibfield  {journal}
  {\bibinfo  {journal} {Nat. Photonics}\ }\textbf {\bibinfo {volume} {16}},\
  \bibinfo {pages} {59} (\bibinfo {year} {2022})}\BibitemShut {NoStop}%
\bibitem [{\citenamefont {Berkeland}\ \emph {et~al.}(1998)\citenamefont
  {Berkeland}, \citenamefont {Miller}, \citenamefont {Bergquist}, \citenamefont
  {Itano},\ and\ \citenamefont {Wineland}}]{Berkeland1998}%
  \BibitemOpen
  \bibfield  {author} {\bibinfo {author} {\bibfnamefont {D.}~\bibnamefont
  {Berkeland}}, \bibinfo {author} {\bibfnamefont {J.}~\bibnamefont {Miller}},
  \bibinfo {author} {\bibfnamefont {J.~C.}\ \bibnamefont {Bergquist}}, \bibinfo
  {author} {\bibfnamefont {W.~M.}\ \bibnamefont {Itano}}, \ and\ \bibinfo
  {author} {\bibfnamefont {D.~J.}\ \bibnamefont {Wineland}},\ }\href@noop {}
  {\bibfield  {journal} {\bibinfo  {journal} {J. Appl. Phys.}\ }\textbf
  {\bibinfo {volume} {83}},\ \bibinfo {pages} {5025} (\bibinfo {year}
  {1998})}\BibitemShut {NoStop}%
\bibitem [{\citenamefont {Leibfried}(1999)}]{Leibfried1999}%
  \BibitemOpen
  \bibfield  {author} {\bibinfo {author} {\bibfnamefont {D.}~\bibnamefont
  {Leibfried}},\ }\href {\doibase 10.1103/PhysRevA.60.R3335} {\bibfield
  {journal} {\bibinfo  {journal} {Phys. Rev. A}\ }\textbf {\bibinfo {volume}
  {60}},\ \bibinfo {pages} {R3335} (\bibinfo {year} {1999})}\BibitemShut
  {NoStop}%
\bibitem [{\citenamefont {Warring}\ \emph
  {et~al.}(2013{\natexlab{b}})\citenamefont {Warring}, \citenamefont
  {Ospelkaus}, \citenamefont {Colombe}, \citenamefont {J{\"o}rdens},
  \citenamefont {Leibfried},\ and\ \citenamefont {Wineland}}]{Warring2013a}%
  \BibitemOpen
  \bibfield  {author} {\bibinfo {author} {\bibfnamefont {U.}~\bibnamefont
  {Warring}}, \bibinfo {author} {\bibfnamefont {C.}~\bibnamefont {Ospelkaus}},
  \bibinfo {author} {\bibfnamefont {Y.}~\bibnamefont {Colombe}}, \bibinfo
  {author} {\bibfnamefont {R.}~\bibnamefont {J{\"o}rdens}}, \bibinfo {author}
  {\bibfnamefont {D.}~\bibnamefont {Leibfried}}, \ and\ \bibinfo {author}
  {\bibfnamefont {D.~J.}\ \bibnamefont {Wineland}},\ }\href {\doibase
  10.1103/PhysRevLett.110.173002} {\bibfield  {journal} {\bibinfo  {journal}
  {Phys. Rev. Lett.}\ }\textbf {\bibinfo {volume} {110}},\ \bibinfo {pages}
  {173002} (\bibinfo {year} {2013}{\natexlab{b}})}\BibitemShut {NoStop}%
\bibitem [{\citenamefont {Sutherland}\ \emph {et~al.}(2022)\citenamefont
  {Sutherland}, \citenamefont {Srinivas},\ and\ \citenamefont
  {Allcock}}]{Sutherland2022}%
  \BibitemOpen
  \bibfield  {author} {\bibinfo {author} {\bibfnamefont {R.}~\bibnamefont
  {Sutherland}}, \bibinfo {author} {\bibfnamefont {R.}~\bibnamefont
  {Srinivas}}, \ and\ \bibinfo {author} {\bibfnamefont {D.~T.~C.}\ \bibnamefont
  {Allcock}},\ }\href@noop {} {\bibfield  {journal} {\bibinfo  {journal}
  {arXiv:2206.06546}\ } (\bibinfo {year} {2022})}\BibitemShut {NoStop}%
\bibitem [{\citenamefont {Rabi}(1937)}]{Rabi1937}%
  \BibitemOpen
  \bibfield  {author} {\bibinfo {author} {\bibfnamefont {I.~I.}\ \bibnamefont
  {Rabi}},\ }\href {\doibase 10.1103/PhysRev.51.652} {\bibfield  {journal}
  {\bibinfo  {journal} {Phys. Rev.}\ }\textbf {\bibinfo {volume} {51}},\
  \bibinfo {pages} {652} (\bibinfo {year} {1937})}\BibitemShut {NoStop}%
\bibitem [{\citenamefont {Srinivas}(2020)}]{Srinivas2020}%
  \BibitemOpen
  \bibfield  {author} {\bibinfo {author} {\bibfnamefont {R.}~\bibnamefont
  {Srinivas}},\ }\emph {\bibinfo {title} {Laser-free trapped-ion quantum logic
  with a radiofrequency magnetic field gradient}},\ \href@noop {} {Ph.D.
  thesis},\ \bibinfo  {school} {University of Colorado, Boulder} (\bibinfo
  {year} {2020})\BibitemShut {NoStop}%
\bibitem [{sup()}]{supplementary}%
  \BibitemOpen
  \href@noop {} {}\bibinfo {note} {See Supplemental Material.}\BibitemShut
  {Stop}%
\bibitem [{\citenamefont {Chou}\ \emph {et~al.}(2017)\citenamefont {Chou},
  \citenamefont {Kurz}, \citenamefont {Hume}, \citenamefont {Plessow},
  \citenamefont {Leibrandt},\ and\ \citenamefont {Leibfried}}]{Chou2017}%
  \BibitemOpen
  \bibfield  {author} {\bibinfo {author} {\bibfnamefont {C.-W.}\ \bibnamefont
  {Chou}}, \bibinfo {author} {\bibfnamefont {C.}~\bibnamefont {Kurz}}, \bibinfo
  {author} {\bibfnamefont {D.~B.}\ \bibnamefont {Hume}}, \bibinfo {author}
  {\bibfnamefont {P.~N.}\ \bibnamefont {Plessow}}, \bibinfo {author}
  {\bibfnamefont {D.~R.}\ \bibnamefont {Leibrandt}}, \ and\ \bibinfo {author}
  {\bibfnamefont {D.}~\bibnamefont {Leibfried}},\ }\href@noop {} {\bibfield
  {journal} {\bibinfo  {journal} {Nature}\ }\textbf {\bibinfo {volume} {545}},\
  \bibinfo {pages} {203} (\bibinfo {year} {2017})}\BibitemShut {NoStop}%
\bibitem [{\citenamefont {Kozlov}\ \emph {et~al.}(2018)\citenamefont {Kozlov},
  \citenamefont {Safronova}, \citenamefont {Crespo L\'opez-Urrutia},\ and\
  \citenamefont {Schmidt}}]{Kozlov2018}%
  \BibitemOpen
  \bibfield  {author} {\bibinfo {author} {\bibfnamefont {M.~G.}\ \bibnamefont
  {Kozlov}}, \bibinfo {author} {\bibfnamefont {M.~S.}\ \bibnamefont
  {Safronova}}, \bibinfo {author} {\bibfnamefont {J.~R.}\ \bibnamefont {Crespo
  L\'opez-Urrutia}}, \ and\ \bibinfo {author} {\bibfnamefont {P.~O.}\
  \bibnamefont {Schmidt}},\ }\href {\doibase 10.1103/RevModPhys.90.045005}
  {\bibfield  {journal} {\bibinfo  {journal} {Rev. Mod. Phys.}\ }\textbf
  {\bibinfo {volume} {90}},\ \bibinfo {pages} {045005} (\bibinfo {year}
  {2018})}\BibitemShut {NoStop}%
\bibitem [{\citenamefont {Micke}\ \emph {et~al.}(2020)\citenamefont {Micke},
  \citenamefont {Leopold}, \citenamefont {King}, \citenamefont {Benkler},
  \citenamefont {Spie{\ss}}, \citenamefont {Schm{\" o}ger}, \citenamefont
  {Schwarz}, \citenamefont {Crespo L{\'o}pez-Urrutia},\ and\ \citenamefont
  {Schmidt}}]{Micke2020}%
  \BibitemOpen
  \bibfield  {author} {\bibinfo {author} {\bibfnamefont {P.}~\bibnamefont
  {Micke}}, \bibinfo {author} {\bibfnamefont {T.}~\bibnamefont {Leopold}},
  \bibinfo {author} {\bibfnamefont {S.~A.}\ \bibnamefont {King}}, \bibinfo
  {author} {\bibfnamefont {E.}~\bibnamefont {Benkler}}, \bibinfo {author}
  {\bibfnamefont {L.~J.}\ \bibnamefont {Spie{\ss}}}, \bibinfo {author}
  {\bibfnamefont {L.}~\bibnamefont {Schm{\" o}ger}}, \bibinfo {author}
  {\bibfnamefont {M.}~\bibnamefont {Schwarz}}, \bibinfo {author} {\bibfnamefont
  {J.~R.}\ \bibnamefont {Crespo L{\'o}pez-Urrutia}}, \ and\ \bibinfo {author}
  {\bibfnamefont {P.~O.}\ \bibnamefont {Schmidt}},\ }\href@noop {} {\bibfield
  {journal} {\bibinfo  {journal} {Nature}\ }\textbf {\bibinfo {volume} {578}},\
  \bibinfo {pages} {60} (\bibinfo {year} {2020})}\BibitemShut {NoStop}%
\bibitem [{\citenamefont {Matthiesen}\ \emph {et~al.}(2021)\citenamefont
  {Matthiesen}, \citenamefont {Yu}, \citenamefont {Guo}, \citenamefont
  {Alonso},\ and\ \citenamefont {H\"affner}}]{Matthiesen2021}%
  \BibitemOpen
  \bibfield  {author} {\bibinfo {author} {\bibfnamefont {C.}~\bibnamefont
  {Matthiesen}}, \bibinfo {author} {\bibfnamefont {Q.}~\bibnamefont {Yu}},
  \bibinfo {author} {\bibfnamefont {J.}~\bibnamefont {Guo}}, \bibinfo {author}
  {\bibfnamefont {A.~M.}\ \bibnamefont {Alonso}}, \ and\ \bibinfo {author}
  {\bibfnamefont {H.}~\bibnamefont {H\"affner}},\ }\href {\doibase
  10.1103/PhysRevX.11.011019} {\bibfield  {journal} {\bibinfo  {journal} {Phys.
  Rev. X}\ }\textbf {\bibinfo {volume} {11}},\ \bibinfo {pages} {011019}
  (\bibinfo {year} {2021})}\BibitemShut {NoStop}%
\bibitem [{\citenamefont {Leibfried}\ \emph {et~al.}(2007)\citenamefont
  {Leibfried}, \citenamefont {Knill}, \citenamefont {Ospelkaus},\ and\
  \citenamefont {Wineland}}]{Leibfried2007}%
  \BibitemOpen
  \bibfield  {author} {\bibinfo {author} {\bibfnamefont {D.}~\bibnamefont
  {Leibfried}}, \bibinfo {author} {\bibfnamefont {E.}~\bibnamefont {Knill}},
  \bibinfo {author} {\bibfnamefont {C.}~\bibnamefont {Ospelkaus}}, \ and\
  \bibinfo {author} {\bibfnamefont {D.~J.}\ \bibnamefont {Wineland}},\ }\href
  {\doibase 10.1103/PhysRevA.76.032324} {\bibfield  {journal} {\bibinfo
  {journal} {Phys. Rev. A}\ }\textbf {\bibinfo {volume} {76}},\ \bibinfo
  {pages} {032324} (\bibinfo {year} {2007})}\BibitemShut {NoStop}%
\bibitem [{\citenamefont {Bowler}\ \emph {et~al.}(2012)\citenamefont {Bowler},
  \citenamefont {Gaebler}, \citenamefont {Lin}, \citenamefont {Tan},
  \citenamefont {Hanneke}, \citenamefont {Jost}, \citenamefont {Home},
  \citenamefont {Leibfried},\ and\ \citenamefont {Wineland}}]{Bowler2012}%
  \BibitemOpen
  \bibfield  {author} {\bibinfo {author} {\bibfnamefont {R.}~\bibnamefont
  {Bowler}}, \bibinfo {author} {\bibfnamefont {J.}~\bibnamefont {Gaebler}},
  \bibinfo {author} {\bibfnamefont {Y.}~\bibnamefont {Lin}}, \bibinfo {author}
  {\bibfnamefont {T.~R.}\ \bibnamefont {Tan}}, \bibinfo {author} {\bibfnamefont
  {D.}~\bibnamefont {Hanneke}}, \bibinfo {author} {\bibfnamefont {J.~D.}\
  \bibnamefont {Jost}}, \bibinfo {author} {\bibfnamefont {J.~P.}\ \bibnamefont
  {Home}}, \bibinfo {author} {\bibfnamefont {D.}~\bibnamefont {Leibfried}}, \
  and\ \bibinfo {author} {\bibfnamefont {D.~J.}\ \bibnamefont {Wineland}},\
  }\href {\doibase 10.1103/PhysRevLett.109.080502} {\bibfield  {journal}
  {\bibinfo  {journal} {Phys. Rev. Lett.}\ }\textbf {\bibinfo {volume} {109}},\
  \bibinfo {pages} {080502} (\bibinfo {year} {2012})}\BibitemShut {NoStop}%
\bibitem [{\citenamefont {Pino}\ \emph {et~al.}(2021)\citenamefont {Pino},
  \citenamefont {Dreiling}, \citenamefont {Figgatt}, \citenamefont {Gaebler},
  \citenamefont {Moses}, \citenamefont {Allman}, \citenamefont {Baldwin},
  \citenamefont {Foss-Feig}, \citenamefont {Hayes}, \citenamefont {Mayer},
  \citenamefont {Ryan-Anderson},\ and\ \citenamefont {Neyenhuis}}]{Pino2020}%
  \BibitemOpen
  \bibfield  {author} {\bibinfo {author} {\bibfnamefont {J.~M.}\ \bibnamefont
  {Pino}}, \bibinfo {author} {\bibfnamefont {J.~M.}\ \bibnamefont {Dreiling}},
  \bibinfo {author} {\bibfnamefont {C.}~\bibnamefont {Figgatt}}, \bibinfo
  {author} {\bibfnamefont {J.~P.}\ \bibnamefont {Gaebler}}, \bibinfo {author}
  {\bibfnamefont {S.~A.}\ \bibnamefont {Moses}}, \bibinfo {author}
  {\bibfnamefont {M.~S.}\ \bibnamefont {Allman}}, \bibinfo {author}
  {\bibfnamefont {C.~H.}\ \bibnamefont {Baldwin}}, \bibinfo {author}
  {\bibfnamefont {M.}~\bibnamefont {Foss-Feig}}, \bibinfo {author}
  {\bibfnamefont {D.}~\bibnamefont {Hayes}}, \bibinfo {author} {\bibfnamefont
  {K.}~\bibnamefont {Mayer}}, \bibinfo {author} {\bibfnamefont
  {C.}~\bibnamefont {Ryan-Anderson}}, \ and\ \bibinfo {author} {\bibfnamefont
  {B.}~\bibnamefont {Neyenhuis}},\ }\href@noop {} {\bibfield  {journal}
  {\bibinfo  {journal} {Nature}\ }\textbf {\bibinfo {volume} {592}},\ \bibinfo
  {pages} {209} (\bibinfo {year} {2021})}\BibitemShut {NoStop}%
\bibitem [{\citenamefont {Johansson}\ \emph {et~al.}(2013)\citenamefont
  {Johansson}, \citenamefont {Nation},\ and\ \citenamefont
  {Nori}}]{Johansson2013}%
  \BibitemOpen
  \bibfield  {author} {\bibinfo {author} {\bibfnamefont {J.~R.}\ \bibnamefont
  {Johansson}}, \bibinfo {author} {\bibfnamefont {P.~D.}\ \bibnamefont
  {Nation}}, \ and\ \bibinfo {author} {\bibfnamefont {F.}~\bibnamefont
  {Nori}},\ }\href@noop {} {\bibfield  {journal} {\bibinfo  {journal} {Comput.
  Phys. Commun.}\ }\textbf {\bibinfo {volume} {184}},\ \bibinfo {pages} {1234}
  (\bibinfo {year} {2013})}\BibitemShut {NoStop}%
\end{thebibliography}%

\clearpage

\section{Supplemental Material}

\setcounter{equation}{0}
\renewcommand{\theequation}{S\arabic{equation}}

\section{Derivation of spin-flip Hamiltonian}

From Eq.~(\ref{eq_bare}), the interaction of a spin-dependent gradient and an electric field is

\begin{align*}
\hat{H} = &\hbar\Omega_g \hat{\sigma}_i\cos{\omega_g t}(\hat{a}+\hat{a}^\dagger) \\ \nonumber
+&\hbar\Omega_e \cos{(\omega_e t+\phi_e)}(\hat{a}+\hat{a}^\dagger),
\end{align*}

\noindent where the first line describes a spin-dependent gradient with Rabi frequency $\Omega_g$ oscillating at a frequency $\omega_g$. The second term corresponds to an electric field with Rabi frequency $\Omega_e$ oscillating at frequency $\omega_e$ and phase $\phi_e$. For simplicity, we set the phase of the gradient to be zero. The gradient couples the internal states of an ion to its motion via the Pauli spin operator $\hat{\sigma}_i$, where $i\in\{x, y, z\}$, and the creation (annihilation) operator $\hat{a}^\dagger$ $(\hat{a})$. The electric field Rabi frequency is $\Omega_e\equiv {qEr_0}/{\hbar}$, where the ion charge is $q$,$E$ is its amplitude along the motional mode, and $\hbar$ is the reduced Planck's constant. The ground state extent of the ion motion is $r_0=\sqrt{\hbar/(2M\omega_m)}$, where $M$ is the ion mass and $\omega_m$ is the motional frequency.

The Hamiltonian describing the ion's qubit and motional frequencies is 

\begin{align}
    \hat{H}_0=\frac{\hbar\omega_0}{2}\hat{\sigma}_z + \hbar\omega_m\hat{a}^\dagger\hat{a}.
\end{align}

\noindent The ion qubit frequency is $\omega_0$. Transforming $\hat{H}$ into the interaction picture with respect to $\hat{H}_0$, for $i=x$, we obtain

\begin{align}
    \hat{H}_g = &\frac{\hbar\Omega_g}{2}(\hat{\sigma}_+ e^{-i\delta t}+\hat{\sigma}_- e^{i\delta t})(\hat{a}e^{-i\omega_m t}+\hat{a}^\dagger e^{i\omega_m t}), \\
    \hat{H}_e = &\frac{\hbar\Omega_e}{2}\Bigl[e^{i(\omega_e t+\phi_e)}+e^{-i(\omega_e t+\phi_e)}\Bigr] (\hat{a}e^{-i\omega_m t}+\hat{a}^\dagger e^{i\omega_m t}).
\end{align}

\noindent We now transform $\hat{H}_g$ into the interaction picture with respect to $\hat{H}_e$. The latter corresponds to the Hamiltonian of a driven harmonic oscillator. Since $\hat{H}_e$ does not commute with itself at all times, we make use of the Magnus expansion to find the corresponding time-evolution operator $\hat{U}_e(t)$. The second-order term of the expansion is proportional to the identity operator, which only contributes a global phase, and thus higher-order terms vanish. Omitting the global phase, we find that $\hat{U}_e(t)$ is given by a displacement operator
\begin{equation}
    \hat{U}_e(t) = \hat{D}\big(\alpha(t)\big) = e^{\alpha(t) \hat{a}^{\dagger} - \alpha^*(t) \hat{a}}
\end{equation}
with

\begin{align}
    \alpha(t) = \frac{\Omega_e}{2} \left[e^{-i\phi_e} \frac{e^{-i(\omega_e-\omega_m)t} -1 }{\omega_e - \omega_m}-e^{i\phi_e} \frac{e^{i(\omega_e+\omega_m)t} - 1}{\omega_e + \omega_m}\right].
\end{align}

\noindent The gradient term $\hat{H}_g$ transforms as ${\hat{H}_I=\hat{U}_e^{\dagger}\hat{H}_g\hat{U}_e}$. Noting that $\hat{U}_e(t)$ only acts on the motional degree of freedom and using $\hat{D}^{\dagger}(\alpha) \hat{a} \hat{D}(\alpha) = \hat{a} + \alpha$ and $\hat{D}^{\dagger}(\alpha) \hat{a}^{\dagger} \hat{D}(\alpha) = \hat{a}^{\dagger} + \alpha^*$, we obtain
\begin{align}
    \hat{H}_I =  &\frac{\hbar\Omega_g}{2}(\hat{\sigma}_+ e^{-i\delta t}+\hat{\sigma}_- e^{i\delta t})(\hat{a}e^{-i\omega_m t}+\hat{a}^\dagger e^{i\omega_m t})\nonumber \\
    +&\frac{\hbar\Omega_g\Omega_e}{4}(\hat{\sigma}_+e^{-i\delta t}+\hat{\sigma}_-e^{i\delta t})\times \nonumber \\
    &\biggl(\Bigl[e^{-i(\omega_e t + \phi_e)}+e^{i(\omega_e t + \phi_e)}\Bigr]\Bigl(\frac{1}{\omega_e-\omega_m}-\frac{1}{\omega_e+\omega_m}\Bigr) \nonumber \\
    +&\frac{e^{i(\omega_m t-\phi_e)} + e^{-i(\omega_m t - \phi_e)}}{\omega_e+\omega_m} - \frac{e^{i(\omega_m t+\phi_e)} + e^{-i(\omega_m t + \phi_e)}}{\omega_e-\omega_m}\biggr).
\end{align}

\noindent The first line corresponds only to $\hat{H}_g$. When $\delta=\pm\omega_e$, we obtain Eq.~(\ref{eq_sf})

\begin{align*}
\hat{H}_\textrm{eff} = \frac{\hbar\Omega_g\Omega_e\omega_m}{2(\omega_e^2-\omega_m^2)}(\cos{\phi_e}\hat{\sigma}_x\mp\sin{\phi_e}\hat{\sigma}_y).
\end{align*}

\noindent Here, we can define the effective Rabi frequency of this interaction as 

\begin{align}
\Omega_\textrm{eff}\equiv\frac{\Omega_g\Omega_e\omega_m}{\omega_e^2-\omega_m^2}.
\end{align}

\section{Estimate of electric field at ion}

We infer the strength of the electric field by applying an oscillating voltage resonant with the ion motion and measuring the coherent displacement of the ion. We first cool the ion close to the ground state of all three motional modes through an initial stage of Doppler cooling, followed by resolved-sideband cooling using the 729~nm and 854~nm lasers. This cooling achieves an average occupation of $\bar{n} \approx 0.01$ for the radial modes and $\bar{n} \approx 0.1$ for the axial mode. We then apply the oscillating voltage for a duration $t$ to the same dc electrode that is used for the experiments presented in the main text. The electric field results in the coherent state $\hat{D}(\alpha)\ket{0}$, where $|\alpha(t)| = q E r_0 / (2 \hbar) \left[i t + \left(e^{i 2 \omega_m t} - 1\right) / \left(2 \omega_m\right)\right]$. We have assumed $\phi_e = 0$ here for convenience since the choice of phase is not relevant for subsequent measurements. 

To estimate the value of $\alpha$, we drive the blue sideband of the mode in question using 729~nm laser pulses of variable duration. We then fit the resulting Rabi flops with a theoretical model to extract $|\alpha|$. We repeat this for different durations $t$. From the measured slope, we can estimate the magnitude of the electric field $E$ along the mode using $d|\alpha(t)| / dt = q E r_0 / (2 \hbar)$, neglecting the small contribution from the term rotating at $2 \omega_m$. We find $E_y = 614(3)\,$\,mV/m and $E_z = 175(1)\,$\,mV/m for the two radial modes, with an AWG output amplitude of 0.1.

\section{Pulse shaping}

We employ pulse shaping of the electric field to minimise any residual excitation of the motional modes. The duration of the pulse shaping required scales as $1/|\omega_e-\omega_m|$. In our demonstration, we conservatively use an approximate $\sin^2$ pulse with a 300\,\textmu s  duration for each of the rising and falling edges.  

\section{Numerical simulations}

We perform numerical simulations of $\hat{H}_g$ and $\hat{H}_e$ using QuTiP~\cite{Johansson2013} to estimate the fidelity of a $\pi$ pulse using the forced motion sideband. We set $\omega_m/2\pi=7$\,MHz, $\Omega_g/2\pi=15$\,kHz, $\omega_e/2\pi=5$\,MHz, and $\Omega_e/2\pi=1\,$MHz. The resulting $\pi$ pulse has a duration of about 114\,\textmu s. We have not included any pulse shaping of the electric field and consider only the first 150 Fock states in our simulation. We find that the error is less than $10^{-4}$ even when the motional mode has an initial thermal occupation of $\bar{n}=1$.

\end{document}